\documentclass[11pt,twoside]{article}
\usepackage{asp2010}

\resetcounters

\begin{document}

\title{Nucleation studies under the conditions of carbon-rich AGB star
  envelopes: TiC} 
\author{A. Beate C. Patzer, Matthias Wendt, Christian Chang and
  Detelv S{\"u}lzle 
\affil{Zentrum f{\"u}r Astronomie und Astrophysik, Technische Universit{\"a}t 
       Berlin, Germany}
}

\begin{abstract}
Many studies of especially dust nucleation in winds of carbon-rich
AGB stars consider primarily carbon as dust forming material. But
dust grains formed in such circumstellar envelopes are rather a
mixture of several chemical elements such as titanium or silicon in
addition to the main component carbon as verified by many
investigations of pre-solar grains enclosed in meteorites, for
example. In this contribution we focus on the study of the
nucleation of titanium carbide particles from the gas phase.
Therefore, the necessary properties of molecular titanium carbide
clusters have been estimated within density functional approaches
and first implications on the homogeneous nucleation of TiC are
studied for conditions being representative for circumstellar dust
shells around carbon-rich AGB stars.
\end{abstract}

\section{Constrains from pre--solar TiC 
grain analysis}

Analyses of meteorites revealed the existence of pre--solar TiC grains
with characteristic isotopic signatures originating from AGB stars as
dust sources (e.g. Bernatowicz et al. 2005, Croat et al. 2005). The
detection of \textit{central} inclusions of such TiC particles in
carbon material implicate titanium carbide condensation
\textit{before} carbon dust formation, thus constraining the dust
condensation sequence. Phase stability lines (S=1) of different
carbon materials have been analysed by Bernatowicz et al. (2005)
especially in view of this sequence. They conclude that TiC before C
dust formation is in principle possible at high temperatures (ca.
1570~K -- 1780~K) but low C/O ratios ($\sim$ C/O $<$ 1.2) depending
only weakly on the pressure. However, these considerations are not
taking any details of the nucleation process of titanium carbide
itself into account.

\section{Molecular properties of small titanium carbide clusters and implications for nucleation studies under the conditions of carbon--rich AGB stars}

The investigation of the properties of molecular clusters is essential
for the understanding of dust nucleation in circumstellar environments
of AGB stars. They can be obtained theoretically by electronic
structure techniques to determine the required data of the
microphysical processes involved, which are often not at hand (see
also Patzer 2007). Here, we focus on the properties of small titanium
carbide clusters of mainly stoichiometric composition, which were
studied within the density functional approach (DFT) (cf. e.g. Paar \&
Yang 1989) at the bpl level of theory in conjunction with the standard
medium sized all--electron split valence 6--31G(d) basis set (Frisch
et al. 1984). This theoretical level of approximation is a reasonable
compromise between computational effort and desired numerical accuracy
(cf. Wendt 2008).

The small titanium carbide cluster systems investigated so far show
structural motives of the bulk TiC lattice (fcc). The calculated
positions of the main active IR modes around 15.7~$\mu$m and $\sim$
21~$\mu$m of these titanium carbide clusters, which are caused by
vibrating carbon atoms in a titanium atom 'grid', are in very good
agreement with the results of the TiC cluster measurements of von
Helden et al. (2000). This does not necessarily imply, that the
'21'~$\mu$m feature is caused by TiC dust particles. Strong objections
are given by e.g. Henning \& Mutschke 2001, Chigai et al. 2003, Li 2003.

The nucleation of titanium carbide in AGB winds has been previously
investigated by e.g. Chigai et al. (1999) using a special key
species concept.  Studying first implications we apply here our
findings to the most simple nucleation process, to determine an upper
limit on the stationary homogeneous TiC nucleation rate (see e.g.
Patzer 2004 for more details on nucleation theory). Therefore, only
the addition of monomers is taken into account and depletion effects
are not yet considered. It turns out, that effective TiC nucleation is
possible at high temperatures (see above), but high supersaturation
ratios are absolutely necessary!

\section{Concluding remarks}

Properties of small titanium carbide clusters have been determined
from elaborate DFT studies. A simple application to the homogeneous
process reveals, that in cool and expanding outflows of carbon--rich
AGB stars effective TiC nucleation is in principle possible at high
temperatures, but highly supersaturated conditions are required.

First implications according to the condensation sequence are
therefore: Is TiC condensation before C dust formation really possible
only at low C/O ratios, especially if high S values are necessary
prerequisites? Can such highly supersaturated conditions only occur in
inhomogeneous AGB winds? Are other different TiC nucleation mechanisms
more important, that operate at (maybe) lower supersaturation?  To
adress these questions Ti$_x$C$_y$ clusters in further, chemically
more complex TiC nucleation studies have to be considered.


\end{document}